\documentstyle[12pt]{article}
\newcommand{\De}{\Delta}

\newcommand{\de}{\delta}

\newcommand{\om}{\omega}
\newcommand{\k}{\kappa}
\newcommand{\bea}{\begin{eqnarray}}
\newcommand{\beq}{\begin{equation}}
\newcommand{\eea}{\end{eqnarray}}
\newcommand{\eeq}{\end{equation}}
\begin{document}
\centerline{\Large  Tilted Rotation and Wobbling Motion in Nuclei}
\centerline{W.D. Heiss$^{1}$ and R.G. Nazmitdinov$^{1,2}$}
\centerline{$^1$ Center for Nonlinear Studies and Department of Physics}
\centerline{University of the Witwatersrand, 2050, Johannesburg, South Africa}
\centerline{$^2$ Bogoliubov Laboratory of Theoretical Physics}
\centerline{Joint Institute for Nuclear Research, 141980 Dubna, Russia}

\vspace{0.3cm}

\begin{abstract}
{\sf The self-consistent harmonic oscillator model including
the three-dimensional cranking term
is extended to describe collective excitations
in the random phase approximation.
It is found that quadrupole collective excitations associated with
wobbling motion in rotating nuclei lead to the appearance
of two-- or three--dimensional rotation.}
\end{abstract}

PACS numbers: 21.60.Jz, 21.60 Ev
\vspace{0.2cm}

Nuclear states grouped into $\Delta I=2$ sequences are remarkably well
described in terms of the principal axis rotation which is the
basis of the cranking model \cite{RS}.
The principal axis cranking rotation
was  intuitively justified by a classical rigid
body rotation which is favorable for a uniform rotation around the long or
the short principal axis. However, if the classical body is not
rigid, it may also uniformly rotate about an axis which does not coincide
with the principal axes of the density distribution.
In fact, more than a century ago  Riemann \cite{Rie}
pointed out that such a situation could occur in
the ellipsoidal self-gravitating fluid.
The physical mechanism behind it is the vortex motion
in a macroscopic system.

Numerous experimental observations implying $\Delta I=1$ sequences
in near spherical nuclei \cite{Cl}   raised the question
about a new type of rotation and its physical nature in a
finite quantum many body system.
For particular nuclei the proton and neutron spin vectors
may be oriented along different axes. Consequently, the total angular momentum
could lie at an angle different from one of the principal axes,
i.e.~at a tilted angle.
It seems shell effects are the driving forces
in this case which can be described within
the Tilted Axis Cranking Model \cite{Fr} which assumes a
two--dimensional rotation.
Other intriguing examples are the rotational bands
observed in  $Hf$, $W$ and $Os$ $(Z=72-76)$ nuclei
with $A\sim 180$ \cite{Wal,Crow,P}.
These bands are characterized by
high K-values, where K is the angular momentum
projection on the body-fixed symmetry axis \cite{BM}.
At finite rotational frequency
a competition could occur between states with
large collective angular momentum
oriented perpendicular to the symmetry axis
and high-K states. Such competition gives rise to
the new backbending phenomenon which is expected to be caused
by the crossing of the ground and tilted
rotational bands \cite{Fr}.
The one--dimensional cranking model based on the signature
concept  (the invariance of the wave function under rotation by
$\pi$ around the rotational axis) fails in this
situation, since signature is not conserved for tilted
rotations. Furthermore,
according to the standard Alaga rules \cite{RS,BM},
the quadrupole B(E2) transitions  to the ground band
are forbidden due to the K selection rules $(\De K>2)$.
However, it is observed that some K-isomer states do decay via
quadrupole transitions to the ground state.
It was suggested that either fluctuations
in the orientation of the angular momentum
or  shape fluctuations \cite{Crow}
or tilting degrees of freedom in the rotation
axis \cite{Wal} are responsible for this phenomenon.
It was also observed that the
$\Delta I=2$ sequence of the rotational states
has been superseded by $\Delta I=1$ transitions in the yrast
rotational band at high rotational frequency.
In the present paper we considerably extend a previous
study \cite{hena} and
demonstrate that tilted rotations naturally occur
beyond a critical rotational frequency
as an instability of collective
vibrations caused by the wobbling motion.

At high spin, the dynamical fluctuations of
the shape and angular momentum vector
can be described by the cranking random
phase approximation (CRPA) as formulated in \cite{Mar,JM}
for a one--dimensional rotation.
This approach is the main theoretical tool used for the analysis
of collective excitations in rotating nuclei
(see \cite{KN,Sh,Nak} and references therein).
To understand the main features of the formation of
tilted bands, we assume
the angular momentum and shape fluctuations
to be dominant. For comparison with experimental data
pairing vibrations \cite{Sh,dan} should be taken into account, but
this degree of freedom does not change our main conclusions;
it involves tedious calculations to be postponed
for future publication.
The CRPA and many RPA calculations for non-rotating nuclei
suffer from the inconsistency between the basis generated by
the mean field and the residual two-body interaction.
To facilitate analytical and numerical results and
to avoid spurious solutions due to such inconsistencies,
we base our analysis upon the three--dimensional harmonic oscillator
model.

We start with the single-particle Routhian
\beq
H_{\Omega}=H_0-\vec \Omega \cdot \vec L\equiv
\sum_{j=1}^N({1\over 2m} \vec  p_j\, ^2+{m\over 2}
 (\omega _x^2x_{j}^2+\omega _y^2y_{j}^2+\omega _z^2z_{j}^2)
-\vec \Omega \cdot \vec l_j) \label{raus} \eeq where the
rotational vector $\vec \Omega$ of the cranking term has the
components $(\Omega_x, \Omega_y, \Omega_z) = \Omega (\sin \theta
\cos \phi,\sin \theta \sin \phi,\cos \theta)$. The Hamiltonian,
Eq.(\ref{raus}), represents the simplest mean field Hamiltonian
which reproduces quite well the main properties of rotating
nuclei \cite{BM,NR}. The spectrum of this Hamiltonian, i.e.~the
eigenmodes, is obtained by solving a third order polynomial in
$E^2$. The normal mode operators $a_k^{\dagger }$ and
$a_k,\,k=1,2,3$ ($[a_k, a_l^{\dagger }]=\delta_{k,l}$) are linear
transformations in the coordinates $q_i$ and momenta
$p_i,\,(i=x,y,z)$ with complex coefficients. In this way we obtain
$H_{\Omega}=\sum_{j=1}^N\sum_{i=1,2,3}E_i (n_i+1/2)_j$ with
$n_i=a_i^{\dagger }a_i$. The normal modes are filled from the
bottom which gives the ground state energy in the rotating frame
$E_{{\it tot}}= E_1 \sum_1 + E_2 \sum_2 +E_3 \sum_3$ where
$\sum_k = <\sum _j^N(n_k+1/2)_j>$ and the configuration is
determined by $\sum_1\leq \sum_2\leq \sum_3$.

The minimization of the Hamiltonian
Eq.(\ref{raus})
with respect to the three frequencies $\omega_i$,
subject to the volume conservation condition
$\omega_x \omega_y \omega_z = \omega_0^3$,
yields the self--consistency relation
\beq
\label{con}
\omega _x^2\langle x^2\rangle=\omega _y^2\langle y^2\rangle
=\omega _z^2\langle z^2\rangle.
\eeq
These energy minima and their corresponding values for $\omega _i$
depend on the rotational vector $\vec \Omega $. For given $\Omega $
we search for tilted solutions by seeking the minimum in
the $\theta -\phi$--plane \cite{hena}. Three major results are reported here:
\begin{enumerate}
\item
     starting from oblate, prolate or tri--axial shapes, for sufficiently high
     values of $\Omega $ a local minimum, which is characterized
     by an oblate shape with the symmetry axis coinciding with the rotational axis,
     ($\theta =90^0$, $\phi=0^0$) is {\sl always} obtained;
\item
     increasing further the value
     of $ \Omega $ this local minimum shifts away from $\theta =90^0$ to
     $\theta <90^0$ and possibly $\phi > 0^0$ thus giving rise to a two-- or
     three--dimensional rotation;
\item
    only when starting from tri--axial, near oblate, nuclei
    is this specific local minimum a global and stable minimum. For prolate
    nuclei, this local minimum is attained only for values of $\Omega $
    so large that other (possibly unphysical) minima
    {\sl with different configurations}
    occur at a considerably lower energy, see Fig.1.
    When starting from the outset with an oblate
    nucleus, the global minimum is obtained for the rotational axis
    perpendicular to the symmetry axis; the specific local minimum referred
    to above is then the second lowest minimum with the {\sl same} configuration.
\end{enumerate}

We first focus our attention to tri--axial near oblate nuclei. Here we encounter two
critical values. At $\Omega^{(1)}_{\rm cr}$ there is a shape transition from tri--axial
to axial (oblate) symmetry. This axial shape remains unchanged for a range
$\Omega^{(1)}_{\rm cr}\leq \Omega \leq \Omega^{(2)}_{\rm cr} $
where $\Omega^{(2)}_{\rm cr}$
signifies the onset of a tilted rotation where the rotational axis no longer coincides
with a principal axis. Note that for $0\leq \Omega \leq \Omega^{(2)}_{\rm cr}$
the rotational
axis is aligned to a principal axis which becomes the symmetry axis for
$\Omega \geq \Omega^{(1)}_{\rm cr}$. At this point there is a confluence in
the energy contours
$E(\omega _y,\omega _z)$ of two minima into one at
$\omega _{\perp}=\omega _y=\omega _z$. In
contrast, at the larger value $\Omega =\Omega^{(2)}_{\rm cr}$ the minimum
at $\theta =90^0,\phi =0^0$ bifurcates into two minima as illustrated in Fig.2.
The two minima occurring at $\theta=75^0$ and $\theta=105^0$ ($\phi=0^0$)
appear at first glance
physically equivalent. However, they are not as can be seen when $\Omega $
is increased further
in which case they move into positions with different values of $\phi$ (and
$\theta$) and only one is the global minimum. In fact, the directions
of $\vec{\Omega}$ play a different role in the different octants \cite{chir}.
We interpret the rotation for
$\Omega^{(1)}_{\rm cr}\leq \Omega \leq \Omega^{(2)}_{\rm cr}$ as a K-isomer.
For the tilted rotation the symmetric shape
no longer prevails.

While these findings are obtained from the numerical minimization
procedure we now further substantiate our analysis by an
analytical procedure. Using the fact that the first critical
rotational frequency is associated with a one-dimensional
rotation, we exploit in Eq.(\ref{con}) that at the transition
point $\omega_y=\omega_z=\omega_{\perp}$ and obtain a third order
equation \beq \label{cub}
u^3-\frac{u}{2}+\frac{1}{2}\frac{r-1}{r+1}=0 \eeq where $u=\Omega
/\omega_{\perp}$ and $r=\sum_3 / \sum_2$. From the discriminant
of Eq.(\ref{cub}) we obtain the critical value $r_{\it
cr}=(\sqrt{27}+\sqrt{2})(\sqrt{27}-\sqrt{2})$. It was shown in
\cite{TA} that a prolate system  becomes eventually oblate with
the rotational axis coinciding with the symmetry axis, if the
initial deformation obeys $r \leq r_{\it cr}$. However, according
to our results, this case is not a global energy minimum, in fact
the global minimum is unphysical. In general, for  $r < r_{\it
cr}$ we obtain three solutions \bea \label{ph1}
&&u=\sqrt{\frac{2}{3}} \cos(\frac{\chi+2\pi n}{3}) \quad n=0,1,2\\
&&\cos \chi= 3\sqrt{\frac{3}{2}}\frac{1-r}{1+r}
\eea
These values of the rotational frequency correspond to three bifurcation
points. Below we demonstrate that one of these solutions
is the critical point $\Omega_{\it cr}^{(1)}$
where the lowest vibrational frequency tends to
zero.

The variations in the one-body potential around the equilibrium
deformation determine the effective quadrupole--quadrupole
interaction \cite{v}. The total Hamiltonian can be presented as

\beq
\label{ham} H = H_{\Omega} - \frac{\k
}{2}\sum_{\mu=-2}^2Q_{\mu}^{\dagger}Q_{\mu} \equiv {\tilde H}-\vec
\Omega \cdot \vec L \eeq

While the mean field Hamiltonian Eq.({\ref{raus}) breaks the
rotational symmetry, the Hamiltonian ${\tilde H}$ in
 the total Hamiltonian Eq.(\ref{ham})
fulfills the commutation rules $\left[ \tilde H,L_i \right] = 0$,
$i=x,y,z$. Note that the self-consistent condition Eq.(\ref{con})
fixes the quadrupole strength constant $\k$ in the RPA
calculations. We solve the RPA equation of motion for the general
coordinates ${\cal X}_{\lambda}$ and momenta ${\cal P}_{\lambda}$
(see for details \cite{KN}) \beq [H,{\cal
X}_{\lambda}]=-i\om_{\lambda}{\cal P}_{\lambda}, \quad [H, {\cal
P}_{\lambda}] = i\om_{\lambda}{\cal X}_{\lambda}, \quad [{\cal
X}_{\lambda}, {\cal P}_{\lambda ^{'}}]= i\de_{\lambda,
\lambda^{'}}\quad . \eeq Here $\omega_{\lambda}$ is the RPA
eigenfrequency in the rotating frame and the associated phonon
$O_{\lambda}^{\dagger }=({\cal X}_{\lambda}- i{\cal
P}_{\lambda})/\sqrt{2}$. In contrast to the CRPA approach, the
phonon is in the present model a superposition of different
signature phonons. The degree of the mixture depends on the
tilted angle: the signature and $|{\rm K}|$ are good quantum
numbers, respectively, for rotations perpendicular and parallel
to the symmetry axis. The non-zero solutions appear in pairs $\pm
\hbar \omega_{\lambda}$, we choose solutions with positive norm.

Let us first  focus on the RPA solution which leads to the
transition from the non-axial to the oblate regime of rotation.
Since at this transition point the minimal solution corresponds
to a one--dimensional rotation ($\vec \Omega\equiv (\Omega, 0,
0)$) around the symmetry axis, the angular momentum becomes a
good quantum number. The RPA states can be characterized by the
projection of the angular momentum because
$[L_x,O_{\lambda}^{\dagger }]=\lambda O_{\lambda}^{\dagger }$.
Consequently, we obtain \beq \label{man} [H,O_{\lambda}^{\dagger
}]=[{\tilde H}-\Omega L_x, O_{\lambda}^{\dagger }]= ({\tilde
\omega}_{\lambda}-\lambda \Omega) O_{\lambda}^{\dagger } \equiv
\omega_{\lambda}O_{\lambda}^{\dagger } \eeq From Eq.(\ref{man})
it follows that at the rotational frequency $\Omega_{\it
cr}={\tilde \omega}_{\lambda}/\lambda$ one of the
 RPA frequency should be equal zero.
The solution of the RPA equation for the positive signature
quadrupole phonons with the largest projection $\lambda=-2$ gives
\beq \label{bif}
\omega_{\lambda=-2}=-\omega_{\perp}\sqrt{\frac{2}{3}} (\cos
\frac{\chi+\pi}{3} -\sqrt{3}\sin \frac{\chi+\pi}{3})+2\Omega.
\eeq This mode corresponds to the quadrupole de--excitation which
leads to the state with two units angular momentum less then the
vacuum state (K-isomer state). From Eq.(\ref{bif}) we obtain the
first critical value of the rotational frequency at which the
transition from the non-collective rotation to the non-axial
collective rotation takes place

\beq \label{cr1}
\Omega_{\rm_cr}^{(1)}=\frac{\omega_{\perp}}{2}\sqrt{\frac{2}{3}}(\cos
\frac{\chi+\pi}{3} - \sqrt{3}\sin \frac{\chi+\pi}{3}) \eeq

Therefore, the positive signature quadrupole de-excitation leads
from the K-isomer state to the yrast state with non-axial
quadrupole shape.

Quantum excitations describing the wobbling motion correspond
to the negative signature quadrupole phonons\cite{Mar,JM,Jan}.
These excitations are connected to the yrast line by means of the
quadrupole transitions which carry one unit of the angular
momentum. Similar to the positive signature phonons, the negative
signature phonons  can have zero excitation energy
at particular rotational frequency.
Solution of the RPA equation for the negative signature phonons
with $\lambda=-1$ for the oblate rotation regime leads to the result
\bea
\label{wob}
&&\omega_{\lambda=-1}=-\sqrt{\frac{\omega_x^2+\omega_{\perp}^2}{3}}
(\cos \frac{\psi}{3} -\sqrt{3}\sin \frac{\psi}{3})+\Omega\\
&&\cos \psi= 3\sqrt{3}
\frac{\omega_x^2\omega_{\perp}}{(\omega_x^2+\omega_{\perp}^2)^{3/2}}
\frac{r-1}{1+r}
\eea
The condition $\omega_{\lambda=-1}=0$ yields the second critical
frequency
\beq
\label{cr2}
\Omega_{\rm cr}^{(2)}= \sqrt{\frac{\omega_x^2+\omega_{\perp}^2}{3}}
(\cos \frac{\psi}{3} -\sqrt{3}\sin \frac{\psi}{3}).
\eeq
 Beyond this rotational frequency
the mean field solution corresponds to the stable tilted rotation. We recall that
the expressions given in Eqs.(\ref{cr1}) and (\ref{cr2}) coincide exactly
with the transitional points found in the numerical minimization procedure.
Note that, in contrast to the familiar
phase transition from spherical to  deformed shape
owing to the variation of a strength parameter \cite{RS},
it is here the rotation that leads to a phase transition from tri--axial
to oblate shape and then to the tilted rotation associated again with a
non--axial shape for the same strength parameter of the
residual quadrupole-quadrupole interaction.

According to our analysis, only
nuclei near to oblate shape, when non--rotational, exhibit at a certain rotational frequency
tilted rotation which, for lesser rotational speed, leads to the high K-isomer states.
These states decay through non-axial shapes to the ground states. Nuclei which are
oblate when non--rotational also can have this behavior which corresponds to the second (local)
minimum discussed above. However, they also have states of good signature with an even lower energy
for the same rotational speed and the same configuration. In contrast, for prolate nuclei
the local minimum with these characteristics occurs only at a value $\Omega $ so large that
the global minimum occurs for an absurdly unphysical configuration; for this reason it is
doubtful to attach particular physical significance to the local minimum.

\vskip 1cm

R.G.N. acknowledges financial support from the Foundation for
Research Development of South Africa which was provided under the
auspices of the Russian/South African Agreement on Science and
Technology. He is also thankful for the warm hospitality which he
received from the Department of Physics during his visit
to South Africa. W.D.H. is grateful for the congenial hospitality
at the JINR. This project has been supported in part
by the RFFI under the Grant
00-02-17194.

\bigskip

\centerline{Figure Captions}
\medskip
{\bf Fig.1} Energy contours in the $\omega_x-\omega_y-$plane for a rotational speed
where the onset of an oblate shape has just taken place and corresponds to the local
minimum shown in the top. However, for the nucleus considered ($N=44$, tri--axial near prolate)
a much deeper minimum has developed as an instability as seen in the top right corner of
the bottom picture. The section of the top illustration is indicated in the bottom part.

\medskip
{\bf Fig.2} Energy minimum in the $\phi -\theta $--plane just before (top) and just after
(bottom) the onset of tilted rotation. Note the local maximum developing at the centre.
The example chosen is $N=36$, a tri--axial near oblate nucleus.


\begin{thebibliography}{99}
\bibitem{RS}
             P.~Ring and P.~Schuck, {\it The Nuclear Many-Body Problem}
             (Springer, Berlin, 1980).

\bibitem{Rie}
              B.~Riemann, Abh. K\"on. Ges. Wiss. (G\"ottingen), {\bf 9},1
                         (1860).
\bibitem{Cl}
              R.M.~Clark {\it et al}, Phys.Rev.Lett. {\bf 82}, 3220 (1999);
              H.~Schnare {\it et al}, Phys.Rev.Lett. {\bf 82}, 4408 (1999);
              D.G.~Jenkins {\it et al}, Phys.Rev.Lett. {\bf 83}, 500 (1999);
              D.G.~Jenkins {\it et al}, Phys.Rev. {\bf C58}, 2703 (1998);
              O.~Vogel {\it et al}, Phys.Rev. {\bf C56}, 1338 (1998);
              R.M.~Clark, J.Phys.G: Nucl.Phys. {\bf 25}, 695 (1999).
\bibitem{Fr}
             S.~Frauendorf, Nucl.Phys. {\bf A557}, 259c (1993);
                           Rev.Mod.Phys., submitted.
\bibitem{Wal}  P.M.~Walker {\it et al}, Phys.Lett. {\bf B309}, 14 (1993).
\bibitem{Crow} B.~Crowell {\it et al}, Phys.Rev.Lett. {\bf 72}, 1164 (1994).
\bibitem{P}   T.~Kutsarova {\it et al}, Nucl.Phys. {\bf A587}, 111 (1995);
              N.L.~Gj{\o}rup {\it et al}, Nucl.Phys. {\bf A582}, 369 (1995);
              C.J.~Pearson {\it et al}, Phys.Rev.Lett. {\bf 79}, 605 (1997);
              C.S.~Purry {\it et al}, Nucl.Phys. {\bf A632}, 229 (1998).
\bibitem{BM}
             A.~Bohr and B.R.~Mottelson, {\it Nuclear Structure} Vol. II ,
             ( Benjamin, New York, 1975).

\bibitem{hena} W.D.~Heiss and R.G.~Nazmitdinov, Phys.Lett. {\bf B 397},1 (1997)
\bibitem{Mar}
             E.R.~Marshalek, Nucl.Phys. {\bf A266}, 317 (1976).
\bibitem{JM}
             D.~Janssen and I.N.~Mikhailov, Nucl.Phys. {\bf A318}, 390
             (1979).
\bibitem{KN} J.~Kvasil and R.G.~Nazmitdinov,  Sov.J.Part.Nucl.
             {\bf17}, 265 (1986).
\bibitem{Sh} Y.R.~Shimizu, J.D.~Garrett, R.A.~Broglia,
             M.~Gallardo and E.~Vigezzi, Rev. Mod. Phys. {\bf 61}, 131
             (1989).
\bibitem{Nak} T.~Nakatsukasa, K.~Matsuyanagi, S.~Mizutori and Y.R.~Shimizu,
              Phys.Rev. {\bf C53}, 2213 (1996); G.~Hackman {\it et al},
              Phys.Rev. {\bf C57}, R1056 (1998).

\bibitem{dan}
            F.~D\"onau, D.~Almehed and R.G.~Nazmitdinov,
            Phys.Rev.Lett. {\bf 83}, 280 (1999);
            D.~Almehed,  F.~D\"onau, S.~Frauendorf and R.G.~Nazmitdinov,
            Physica Scripta, in press.
\bibitem{NR} S.G.~Nilsson and I.~Ragnarsson, {\it Shapes and Shells in
              Nuclear Structure} (Cambridge University Press, Cambridge,
              1995).
\bibitem{chir} V.I.~Dimitrov, S.~Frauendorf and F.~D\"onau,
              Phys. Rev. Lett. {\bf  84}, 5732 (2000).
\bibitem{TA} T.~Troudet and R.~Arvieu, Ann.of Phys. (N.Y.) {\bf 134}, 1 (1981).
\bibitem{v} T.~Kishimoto {\it et al}, Phys.Rev.Lett. {\bf 35}, 552 (1975);
            H.~Sakamoto and T.~Kishimoto, Nucl.Phys. {\bf A501}, 205 (1989)
            and references therein.
\bibitem{Jan}
            D.~Janssen, I.N.~Mikhailov, R.G.~Nazmitdinov, B.~Nerlo--Pomorska,
            K.~Pomorski and R.Kh.~Safarov, Phys. Lett. {\bf 79B}, 347 (1978).
\end{thebibliography}
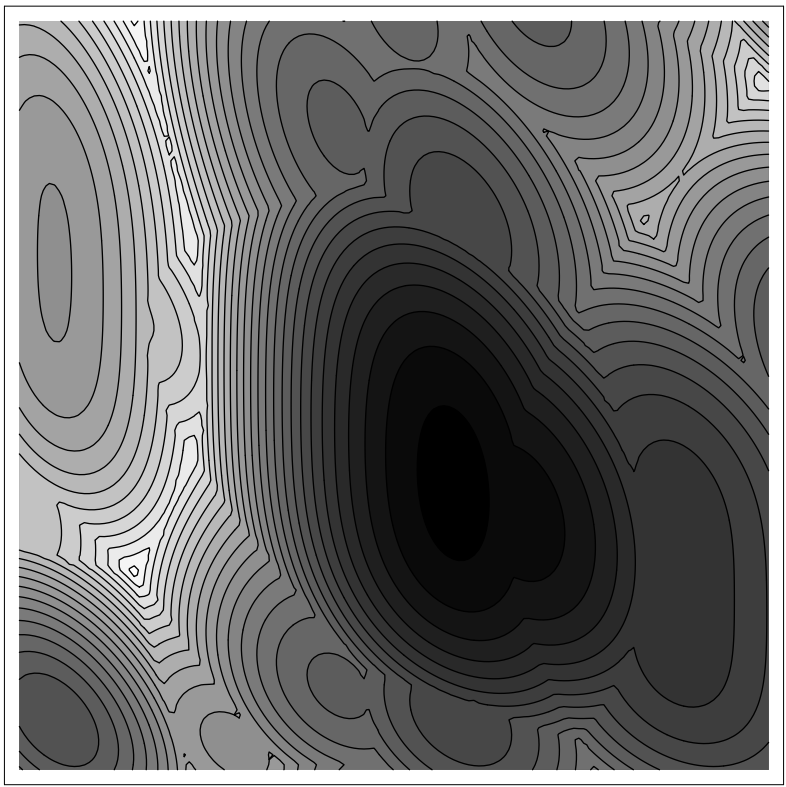
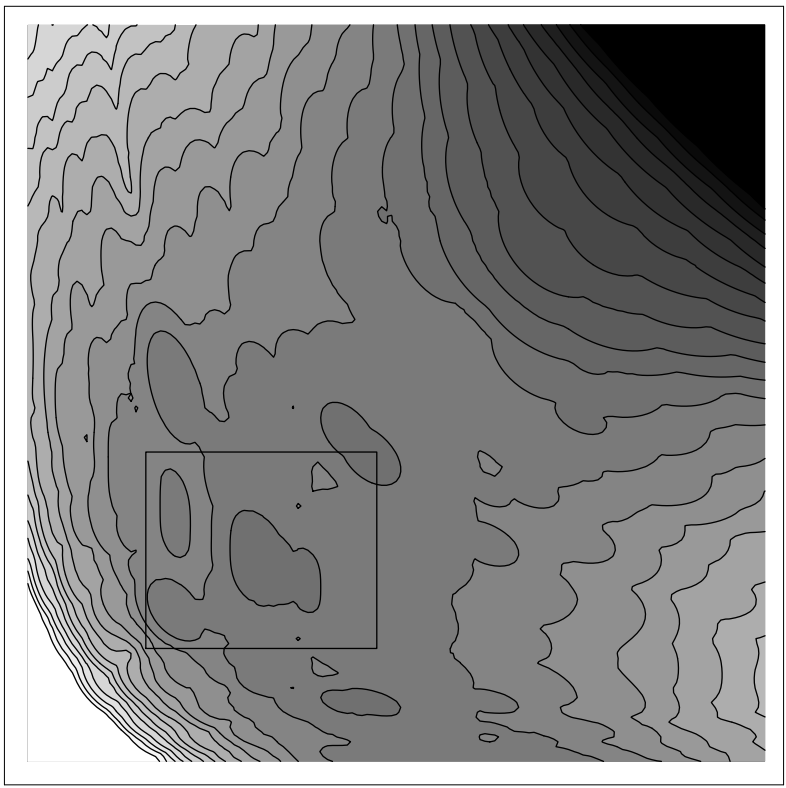
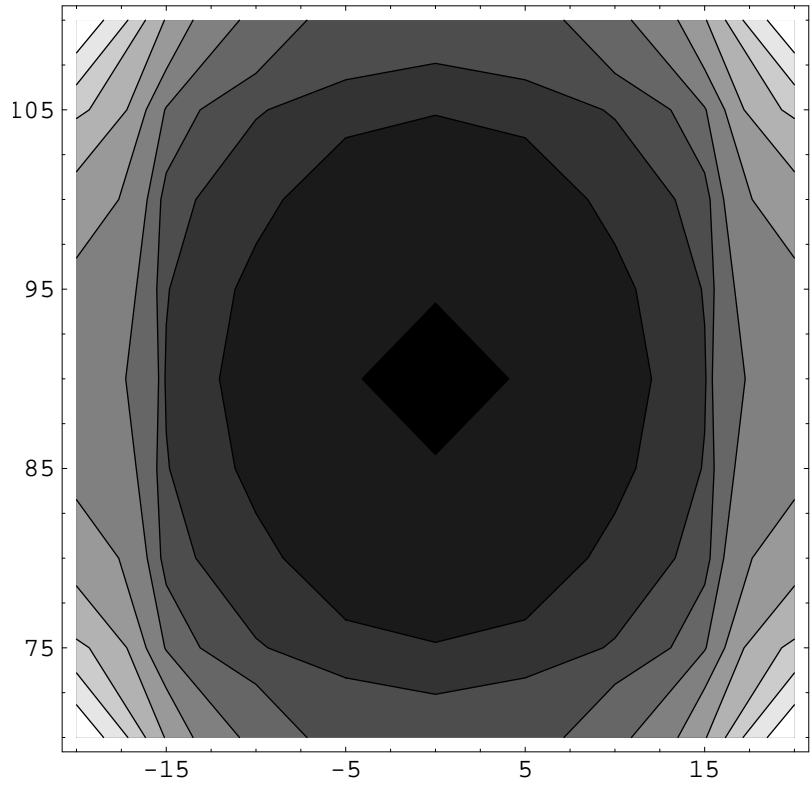
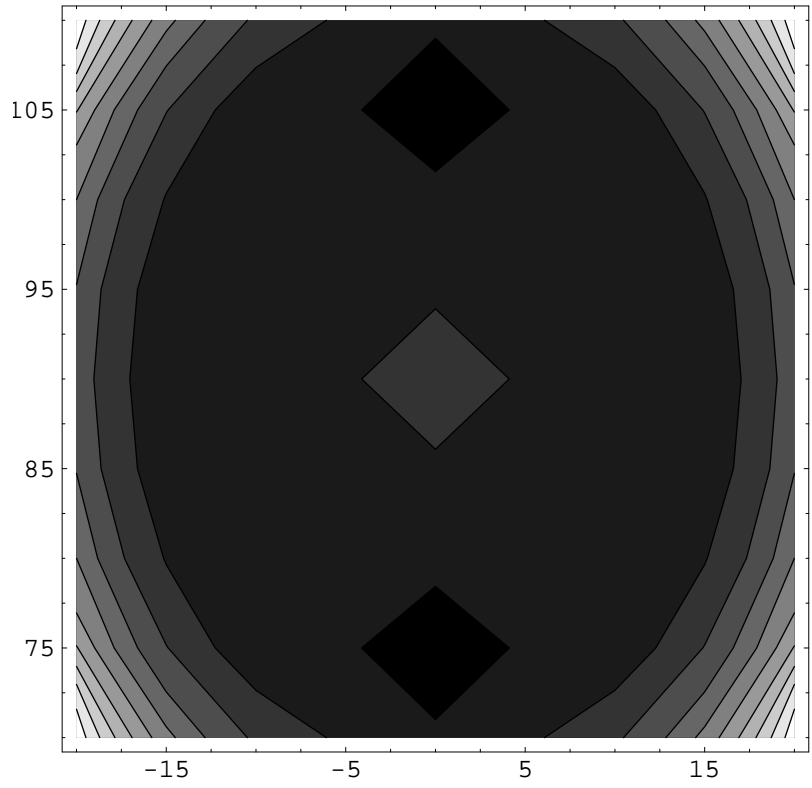
\end{document}